\documentclass[12pt]{article}
\usepackage[top=0.85in,left=2.75in,footskip=0.75in]{geometry}

\usepackage{color,amsmath,amssymb}

\usepackage{changepage}

\usepackage[utf8x]{inputenc}

\usepackage{textcomp,marvosym}

\usepackage{cite}

\usepackage{nameref,hyperref}

\usepackage[right]{lineno}

\usepackage[table]{xcolor}

\usepackage{array}

\newcolumntype{+}{!{\vrule width 2pt}}

\newlength\savedwidth

\raggedright
\usepackage[aboveskip=1pt,labelfont=bf,labelsep=period,justification=raggedright,singlelinecheck=off]{caption}

\makeatletter
\renewcommand{\@biblabel}[1]{\quad#1.}
\makeatother

\date{}

\usepackage{fancyhdr,graphicx}
\usepackage{epstopdf}
\fancyheadoffset[L]{2.25in}
\fancyfootoffset[L]{2.25in}
\begin{document}
\vspace*{0.2in}

\begin{flushleft}
{\Large
\textbf\newline Likelihood analysis of small field polynomial models of inflation yielding a high Tensor-to-Scalar ratio 
}
\newline
\\
Ira Wolfson\textsuperscript{1*},
Ramy Brustein\textsuperscript{1},
\\
\bigskip
\textbf{1} Department of physics, Ben-Gurion University of the Negev, 8410500 Beer-Sheva, Israel
\bigskip

* irawolf@post.bgu.ac.il

\end{flushleft}
\section*{Abstract}
Inflationary potentials, with Planckian field excursions, described by a 6th degree polynomial are studied. We solve the Mukhanov-Sasaki equations exactly and employ a probabilistic approach as well as multinomial fitting to analyse the results. We identify the most likely models which yield a tensor-to-scalar ratio $r=0.01$ in addition to currently allowed Cosmic Microwave Background (CMB) spectrum and observables.
Additionally, we find a significant inter-dependence of CMB observables in these models. This might be an important effect for future analyses, since the different moments of the primordial power spectrum are taken to be independent in the usual Markov chain Monte Carlo methods.

\section*{Author summary}
This work presents the most likely candidates for a small field inflationary potentials, that generate significant GW signal with $r=0.01$, while conforming to known CMB observables. Two methods are used to recover these potentials, one is based on likelihood analysis, while the other is based on multi-variate polynomial fitting. The two methods are in agreement.


\renewcommand{\baselinestretch}{1.5}
\section{Introduction\label{Introduction}}
\begin{figure}[h]
\includegraphics[width=1\textwidth]{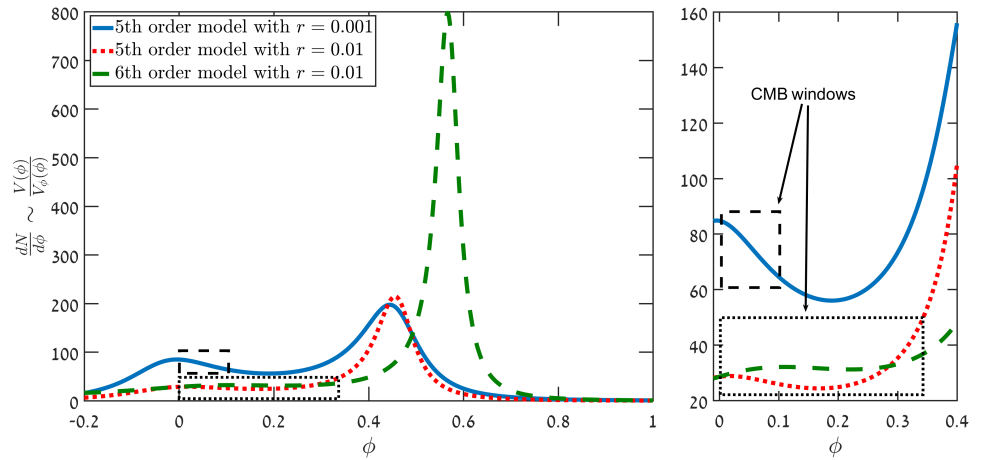}
\caption{A graph depicting $-1/\sqrt{2\epsilon}\sim\frac{V}{V_{\phi}}$ as a function of the inflaton $\phi$ for two fifth-order polynomial models, and a sixth-order polynomial model. For a fifth-order polynomial model with $r=0.001$ (Blue line) the CMB window width is $\sim 8$ e-folds, while the field changes by about $\Delta\phi\sim 0.1$. Most of the e-folds are generated when $\phi$ reaches $\sim 0.4$.  When $r$ is increased the CMB window widens, and approaches the e-fold generating peak (Red dots). While marginally affecting the CMB window width, the introduction of an additional coefficient, $a_6$, allows to shift the peak to higher values of $\phi$, thereby decoupling the CMB window and the e-fold generating peak (Green dash).}
\label{fig:1OverSqrt(2eps)}
\end{figure}

In a previous article \cite{Wolfson:2016vyx}, a class of small field inflationary models which are able to reproduce the currently measured Cosmologic Microwave Background (CMB) observables, while also generating an appreciable primordial Gravitational Wave (GW) signal were studied. The existence of such small field models provides a viable alternative to the large field models that generate a high Tensor-to-Scalar ratio. Our exact analysis was shown to give accurate results \cite{Wolfson:2016vyx}. Models which yield Tensor to Scalar ratio ($r$), of less than $r\lesssim 0.003$ were previously studied in \cite{Wolfson:2016vyx}. The initial study demonstrated a significant difference between analytical Stewart-Lyth \cite{Stewart:1993bc,Lyth:1998xn} estimates and the exact results. This result should be confronted with analyses such as in \cite{Martin:2013tda} where the Stewart-Lyth expression is relied upon, and \cite{Dodelson:2001sh} in which the authors use a Green's function approach and perturbation theory, but assume the log of the input is well behaved. Our method extends and improves the method of the model building technique employed in \cite{BenDayan:2009kv,Hotchkiss:2011gz}. Previous analytical work \cite{Choudhury:2014kma,Chatterjee:2014hna,Choudhury:2015pqa} has shown that a fourth-order polynomial potential is sufficient to generate a high tensor-to-scalar ratio, even up to $r\gtrsim 0.1$. However, it was hard to realize this numerically. It was discovered in \cite{Wolfson:2016vyx}, that a fifth-order polynomial potential was required for generating $0.001\lesssim r\lesssim 0.003$. Furthermore a sixth-order polynomial seems to be required for a tensor-to-scalar ratio greater than $r\gtrsim 0.003$. A simple explanation is offered by observing (see Fig.~\ref{fig:1OverSqrt(2eps)}) that increasing $r$ by factor $\sim 10$, causes the e-folds per field excursion generated at the CMB window to decrease by a factor of $\sim 3$. This means widening the CMB window, and losing the decoupling between the CMB window and the e-fold generating peak. Adding the 6th coefficient pushes the peak from $\phi\in[0.4,0.5]$, to higher values of $\phi$ and decouples these regions.
\subsection{Conventions}
In this article we follow the conventions of \cite{Cabass:2016ldu}. The Primordial Power Spectrum (PPS) is given by:
\begin{align}
	P_{k}=A_{s} \left(\frac{k}{k_{0}}\right)^{n_s-1 +\tfrac{\alpha_s}{2}\log\left(\tfrac{k}{k_{0}}\right) +\tfrac{\beta_s}{6}\left(\log\left(\tfrac{k}{k_{0}}\right)\right)^2}\;.
\end{align}
Our conventions are:
\begin{align}
	n_s-1=\frac{\partial \log P_k}{\partial \log(k)},\\
	\alpha_s=\frac{\partial^{2} \log P_k}{\partial \log(k)^2},\\
	\beta_s=\frac{\partial^{3} \log P_k}{\partial \log(k)^3}.
\end{align}
The PPS is expanded about the pivot scale $k_{0}$. The pivot scale is usually set a-posteriori, at the scale in which the parameters $\{n_s-1,\alpha_s,\beta_s\}$ are minimally dependent \cite{Cortes:2007ak,Liddle:2006ev,Peiris:2006sj}, in Planck + BICEP2 data analyses it is usually set at $k_{0}=0.05\;h Mpc^{-1} $.

\section{Inflationary models}
The small field models previously studied in \cite{Wolfson:2016vyx} yielded results that are consistent with observable data up to values of $r\simeq 0.003$. While these values agree with the current limits on $r$ set by Planck \cite{Ade:2015tva,Ade:2015xua}, we are interested in studying models with higher $r$. The study of small field models is motivated by their appearance in many  fundamental physics frameworks, effective field theory, supergravity \cite{yamaguchi} and string theory \cite{baumann} in successive order of complexity. For models with $r\gtrsim 0.003$, significant running of running is found. This means that while three free parameters (corresponding to $n_{s},\alpha_s,N$) were previously needed, we now need an additional free parameter. Therefore we turn to a model of a degree six polynomial potential. Obviously considering higher degree models complicates the analysis by adding other tunable parameters.
The potential is given by the following polynomial:
\begin{align}
	V=V_{0}\left(1+\sum_{p=1}^{6}a_p \phi^p\right)\;.
\end{align}
It has been shown \cite{BenDayan:2009kv} that the potential can be written as:
\begin{align}
	V=V_0\left(1-\sqrt{\frac{r_0}{8}}\phi +\frac{\eta_0}{2}\phi^2 +\frac{\alpha_0}{3\sqrt{2 r_0}}\phi^3 +a_4\phi^4+a_5\phi^5+a_6\phi^6\right).
\end{align}
However, for simplicity, we express the potential as follows:
\begin{align}
	V=V_0\left(1-\sqrt{\frac{r_0}{8}}\phi +\sum_{p=2}^{6}a_p \phi^p\right)\;,
\end{align}
with the subscript $0$ denoting the value at the CMB point. By setting $\phi_0=0\; ; \;\phi_{end}=1$ we limit ourselves to small field models in which $\Delta\phi=1$ in Planck units, with little effect on CMB observables. 
According to the Lyth bound \cite{Lyth:1996im,Easther:2006qu}, given a Tensor-to-Scalar ratio of $r\simeq 0.01$, the lower bound on the field excursion is approximately given by $\Delta\phi_{4}\gtrsim 0.03\;m_{pl}$. Here $\Delta\phi_{4}$ is the field excursion while the first $\sim 4$ efolds are generated. Our models satisfy this strict bound, as the first 4 efolds or so typically result in $\Delta\phi_4 \sim 0.15$ which is well above $0.03$. The Lyth bound was further extrapolated \cite{Efstathiou:2005tq} to cover the entire inflationary period. Applying this approach to models with $r\sim 0.01$ yields $\Delta\phi\simeq 2\; m_{pl}$. However, in \cite{Hotchkiss:2011gz}, it was shown that in models such as the ones we study, the value of $\Delta\phi$ can be smaller because $\epsilon_H$ is non-monotonic. In this case, $\Delta\phi=1\; m_{pl}$ from the CMB point to the end of inflation is consistent with the Lyth bound.

When the coefficients $\{r_0,a_2,a_3,a_4\}$ are fixed, the remaining coefficients are related by:
\begin{align}
	a_5 =f_1(r_0,a_2,a_3,a_4,a_6),\\
	a_6 = f_2(r_0,a_2,a_3,a_4,N).
\end{align}
The procedure of finding $f_1$ and $f_2$ was explained in detail for the degree 5 polynomial models in \cite{Wolfson:2016vyx}, and here we follow a similar procedure for the degree 6 models. So, ultimately, the model is parametrized by 5 parameters: the two physical parameters $r_0$ and $N$ and the three other parameters ($a_2,a_3,a_4$) that are used to parametrize the $n_{s},\alpha_s,\beta_s$ parameter space. It should be pointed out that $N$ is not an observable, rather $N\sim 50-60$ is a 'soft' constraint. Strictly speaking, $N$ depends on the reheating temperature and only its maximum value can be determined. However for simplicity we treat $N$ as an observable, in order to facilitate the study of a large sample of models.
\section{Coefficient extraction methods\label{methods}}
In this section we explain the two methods for calculating the most likely coefficients $\{a_2,a_3,a_4\}$, given a large number of simulated models and the likelihood data for the CMB observables. This data is available through MCMC analysis of CMB data, such as the Planck data.
\subsection{Likelihood assignment method - Gaussian extraction\label{Probability}}
\begin{figure}[!ht]
\includegraphics[width=1\textwidth]{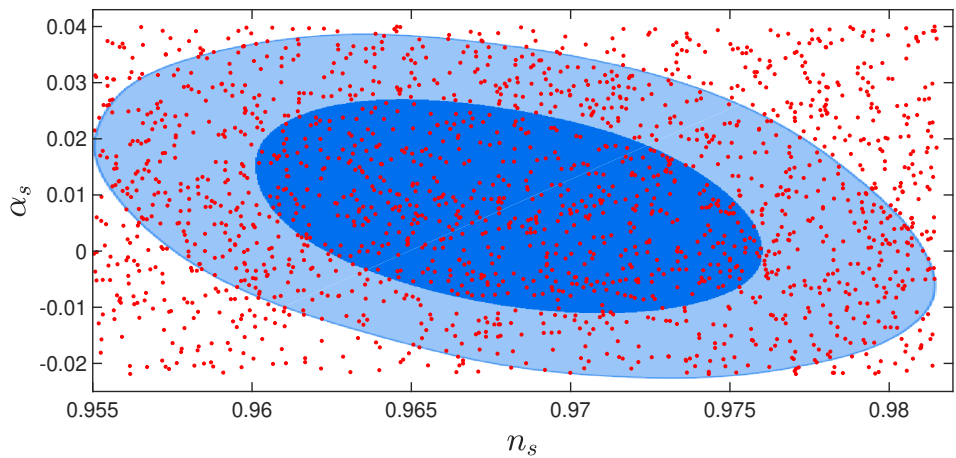}
\caption{Small field inflationary potentials which yield $r=0.01$, as well as PPS observables within 68\% and 99\% confidence levels. Every $(n_s,\alpha_s)$ pair is accessible using these models. The likelihood curves are results of a CosmoMC \cite{Lewis:2002ah} run with the latest BICEP2+Planck \cite{Ade:2015tva} data. \label{main_001} }
\end{figure}
To each potential, after calculating the observables $n_{s},\alpha_s,\beta_s$, we assign a likelihood. For each observable we calculate the likelihood according to the MCMC likelihood analysis of the data sets used. We then assign the product of the likelihoods $L_{(n_s)}\times L_{(\alpha_s)}\times L_{(\beta_s)}$ to the potential. A concrete example is the following:
suppose we extract the trio $(n_{s},\alpha_s,\beta_s)=(0.96,0.011,0.024)$, we look up the likelihoods: $(L_{\left(n_{s}=0.96\right)},L_{\left(\alpha_s=0.011\right)},L_{\left(\beta_s=0.024\right)})$. We now multiply them, and so the likelihood attached to that specific model which yielded these observables is given by $
L_{potential}=L_{\left(n_{s}=0.96\right)}\times L_{\left(\alpha_s=0.011\right)}\times L_{\left(\beta_s=0.024\right)}\;.$ We proceed to extract likelihoods for the different coefficients by a process of marginalization. The expectation is that this method will yield a (roughly) Gaussian distribution for each of the values of $a_2,a_3,a_4$. The advantage of this method is in yielding not only the most likely value, but also the width of the Gaussian. This width can then be used as an indication for the level of tuning that is needed in these models.
\subsubsection{Possible pitfalls}
This method of likelihood assignment is vulnerable in two ways:
\begin{itemize}
	\setlength{\topsep}{0pt}
    \setlength{\itemsep}{0pt}
    \setlength{\parskip}{0pt}
    \setlength{\parsep}{0pt}
\item[a)] To be valid, this method requires a uniform cover of the relevant parameter space, by the potential parameters. If the cover significantly deviates from uniform, the results might be skewed by overweighting areas of negligible weight, or underweighting areas of significant weight. Fig. \ref{main_001} shows a mostly uniform cover.
	\item[b)] Since $ L_{n_s,\alpha_s,\beta_s} \simeq L_{(n_s)}\times L_{(\alpha_s)}\times L_{(\beta_s)}$ only if the paired covariance is small, we must make sure that this is the case. In our underlying MCMC analysis this is indeed the case. The covariance terms are, in general, one to two orders of magnitude smaller than the likelihoods at the tails of the Gaussian.
	\item[c)] We also run the risk of false results if the fit we apply to the data points produced by the numerical analysis yields a large fitting error. However the fitting error of the polynomial function to the $\log PPS - \log k$ data, is usually of the order of $10^{-6}$. This fitting is done over 30 data points generated by the MS equation numerical evaluation, for each potential. The error is calculated as $\Delta=\sqrt{\sum_1^{30}\left((\log PPS)_i-\mathrm{fit}((\log k)_i)\right)^2}$, thus roughly speaking the error per data point is of the order of $10^{-6\sim 7}$. We conclude that the $\log PPS$ function is well fitted.
\end{itemize}
\begin{figure}[!ht]
\includegraphics[width=1\textwidth]{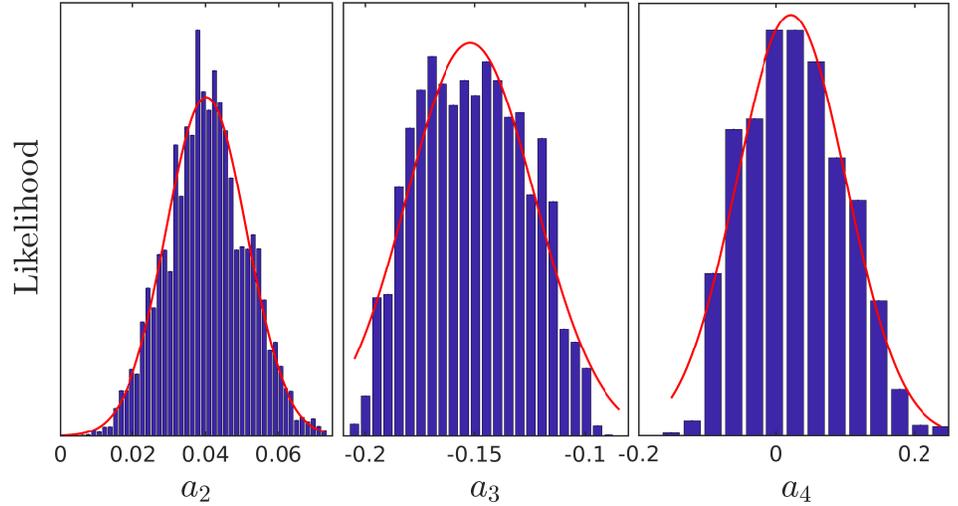}
\caption{The calculated likelihoods for the coefficients $\{a_2$,$a_3$,$a_4\}$, in models with $r=0.01$. The most likely coefficients are given by: $a_2=0.04$, $a_3=-0.15$, $a_4=0.02$. The tuning level for each coefficient is given by Barbieri-Giudice measure \cite{Barbieri:1987fn} and is $(0.375,0.27,5.5)$. \label{Prob_001} }
\end{figure}
\subsection{Multinomial fit\label{multi}}
Another method for calculating the most likely coefficients is by fitting the simulated data with a multinomial function of the CMB observables. We aim to find a set of functions $F_i$ such that, for example, $a_2=F_2(n_{s},\alpha_s,\beta_s)$. We assume that this function is smooth and thus can be expanded in the vicinity of the most likely CMB observables. Hence, we can find a set of multinomials $(F_2,F_3,F_4)$, such that:
\begin{align}
	\begin{array}{ccc}
	a_2&=& F_2(n_{s},\alpha_s,\beta_s)\\
	a_3&=& F_3(n_{s},\alpha_s,\beta_s)\\
	a_4&=& F_4(n_{s},\alpha_s,\beta_s).
	\end{array}
\end{align}
We have found that a quadratic multinomial is sufficiently accurate and that using a higher degree multinomial does not improve the accuracy significantly. Thus we may represent these by a symmetric bilinear form  plus a linear term, as follows:
\begin{align}
	F_i=OB_iO^{\dag} + A_iO^{\dag} +p_{0,i}
\end{align}
where $O=(n_{s},\alpha_s,\beta_s)$, $B_i$ is the bilinear matrix, and the linear coefficient vector is $A_i$.

\subsection{Pivot scale}
So far, we discussed matching potentials and their resulting PPS around the CMB point. However, in order to correctly compare results of the PPS to observables, one has to take into account the pivot scale at which the CMB observables are defined. Since, in this case, the pivot scale is given by $k_0=0.05 \;h Mpc^{-1}$, and the CMB point is at $k\sim 10^{-4}\; hMpc^{-1}$ , the observables in the CMB point and $k_0$ should be related in a simple way only if the spectrum varies slowly with $k$. This is not true for the case at hand.  Two potentials can yield very different power spectra near the CMB point, and nevertheless yield the same observables at the pivot scale. These degeneracies, stem from our limited knowledge of the power spectra on small scales, and at the CMB point. For concreteness take two PPS functions, one that is well approximated by a cubic fit near the pivot scale, and the other that is well approximated only when we consider a quartic fit. Suppose, additionally, that these two PPS functions have the exact same first three coefficients, it follows that they yield the exact same observables $\{n_s,\alpha_s,\beta_s\}$. However if we go to sufficiently small scales, or sufficiently high $k$ values, these functions will diverge. This is also true at the large scale end, where the CMB point is set. Hence the degeneracy.

A possible solution to this problem, is classifying the resulting power spectra by the level of minimal good fit. We define a good fit as one in which the cumulative relative error $\Delta=\sqrt{\sum_k \left(\log(PPS(\log(k)))-fit(\log(k))\right)^2}$, is less than $10^{-7}$.  Given a single power spectrum, we fit our result with a polynomial fit, increasing in order until the accumulated relative error is sufficiently small. The minimal degree polynomial fit that approximates the $\log(PPS)$ function to the aforementioned accuracy is called the minimal good fit.

We then study separately power spectra that are well fitted by cubic polynomials, quartic polynomials etc. In this way we make sure that we compare non-degenerate cases.

\section{Monte Carlo analysis of Cosmic Microwave Background with running of running}
In \cite{Cabass:2016ldu} it was shown that the inclusion of additional parameters, i.e., the running of the spectral index ($\alpha_s$), and the running of the running ($\beta_s$) resolves much of the tension between different data sets. In this section we briefly discuss the effect of considering non-vanishing $\alpha_s$ and $\beta_s$ on the most likely shape of the PPS. First we find $n_s$ when it is the only free parameter. We then use $n_s$, and $\alpha_s$ as the free parameters, and finally we conduct an analysis with $n_s,\alpha_s$, and $\beta_s$ as the free parameters. The shape of the power spectrum changes significantly when running of running is considered.

\begin{figure}[!ht]
\includegraphics[width=1\textwidth]{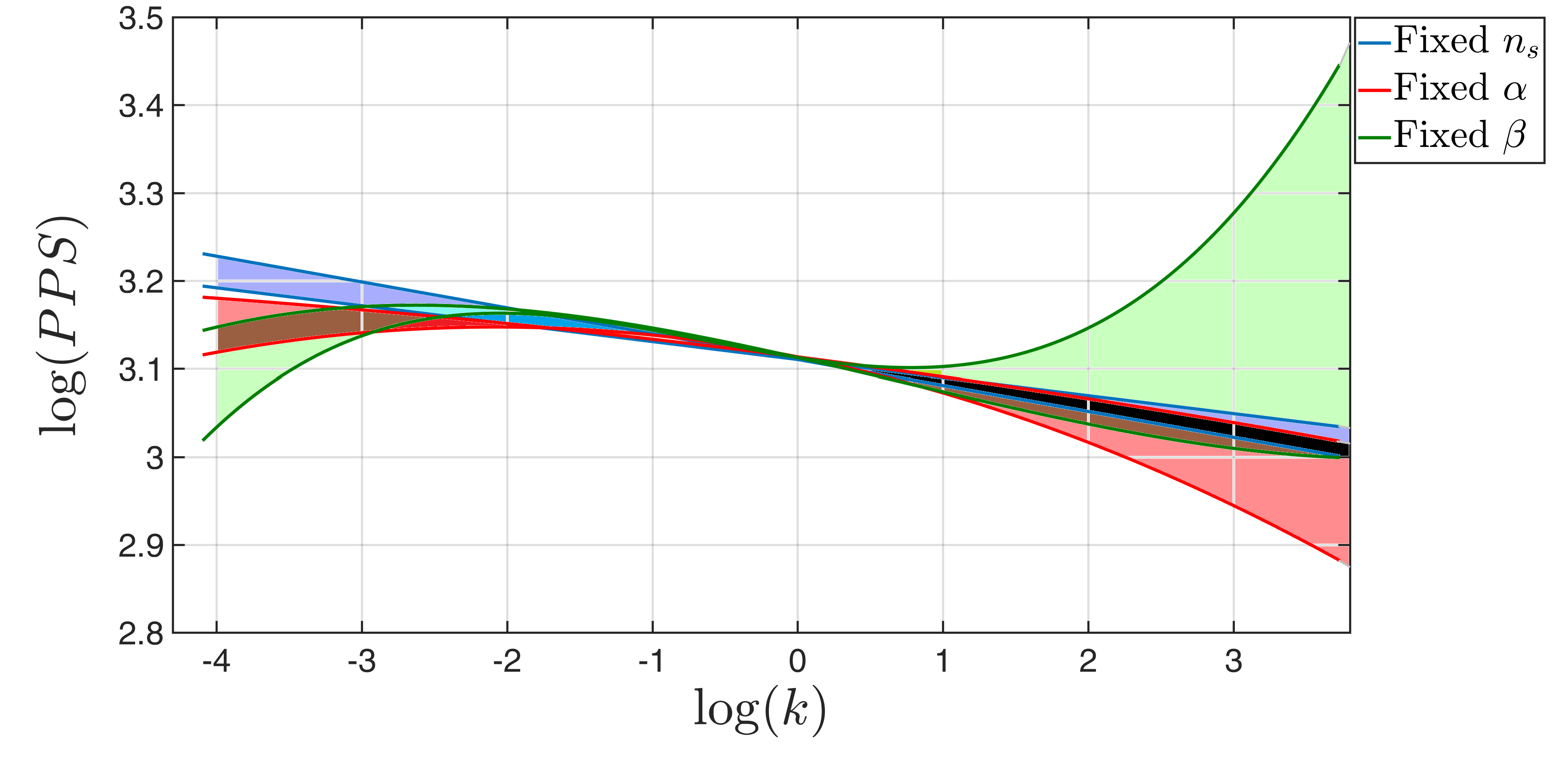}
\caption{Power spectra as recoverd using CosmoMC \cite{Lewis:2002ah} analysis with latest BICEP2+Planck data \cite{Ade:2015tva}. Allowed area (68\% CL) for a fixed $n_s$ analysis is shown (blue). Similarly a fixed $\alpha_s$ analysis (red), and  a fixed $\beta_s$ (green) are shown. The other colores are intersection areas. The pivot scale in this graph is at $\log\left(\frac{k}{k_0}\right)=0$, where $k_0=0.05\;h Mpc^{-1}$. The apparent divergence in high k's is due to the inability of Planck to constrain these k's. This is also shown in Fig. \ref{Power_Spectra_l}. With more data, it will be possible to differentiate between the three possibilities.  \label{POWER_SPECTRA}}
\end{figure}

\begin{figure}[!ht]
\hspace*{-2.3cm}
\includegraphics[height=14cm, width=1.65\textwidth]{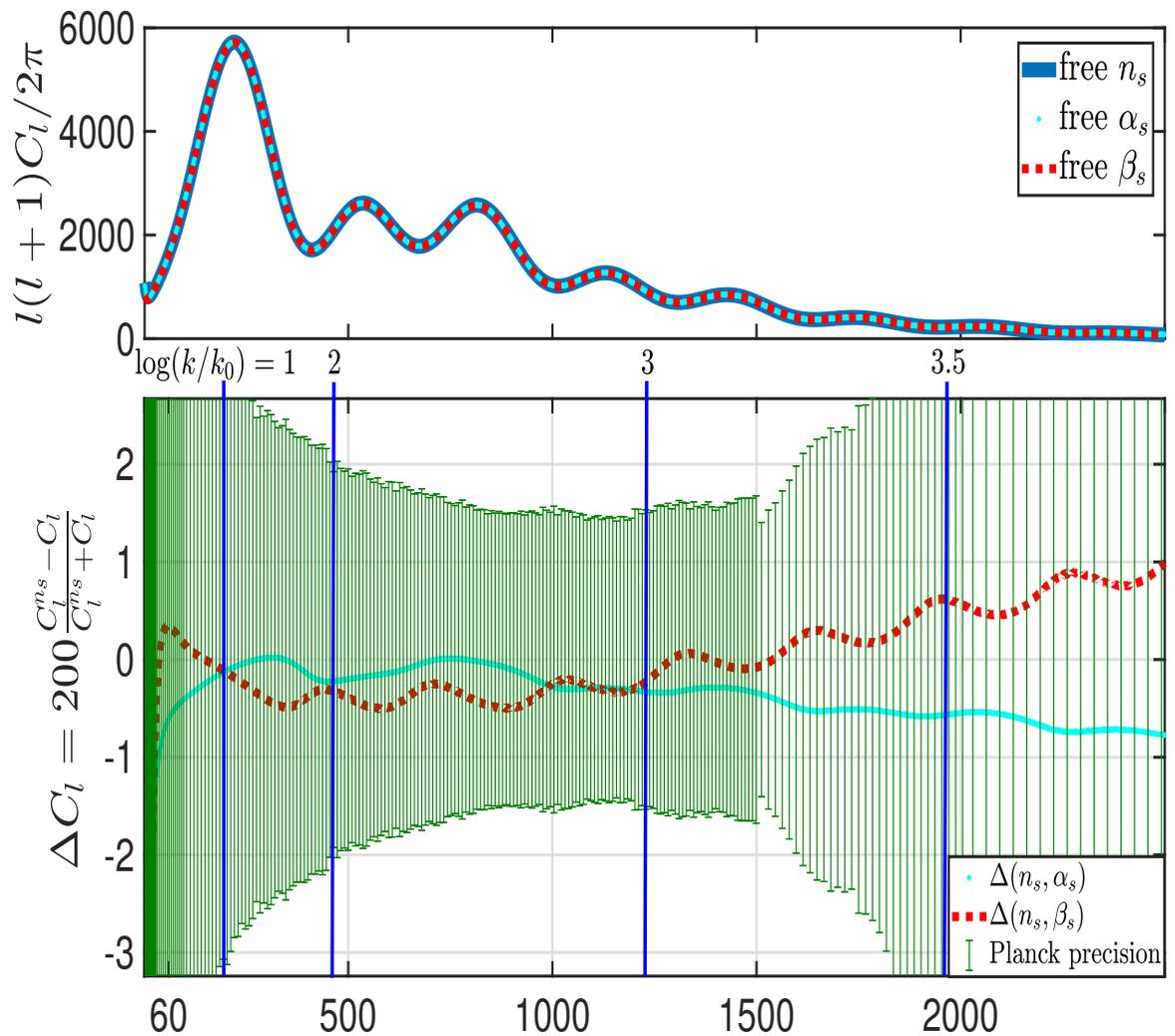}
\caption{Power spectra in the $C_l$'s decomposition (upper panel), with a free $n_s$ (thick blue line), free $\alpha_s$ (thin cyan dots), and free $\beta_s$ (medium red dash). The lower panel shows the realtive difference between the different cases. The relative difference (lower panel) is bound from above by $\sim 1\%$. Additionally the Planck observation error bars are shown.\label{Power_Spectra_l} }
\end{figure}

The data sets that were used are the latest BICEP2+Planck baseline \cite{Ade:2015tva}, along with the low $l$'s \cite{Bennett:2012zja}, low TEB and lensing likelihoods.  The results of these analyses are given in Table \ref{Table_cosmomc}, as well as in Fig. \ref{POWER_SPECTRA}. As expected the resulting power spectra converge at the pivot scale $k_0=0.05\;h Mpc^{-1}$. However for lower $k$'s, the resulting spectra diverge considerably, consistent with cosmic variance. Notably, the spectra also diverge at higher $k$'s. This indicates the inability of current observational data to constrain  the models in this range of $k$'s. This inability is also demonstrated in Fig. \ref{Power_Spectra_l} where, for $l>1500$, the most restrictive data cannot rule out models with significant running, or running of running.
 Figure \ref{Power_Spectra_l} also shows that the three models are virtually indistinguishable in terms of the observed $C_l$'s.

The conclusion is that we will need additional accurate data from smaller cosmic scales to be able to differentiate between the three scenarios. These extra e-folds might come from future missions such as Euclid \cite{Amendola:2016saw}, or from $\mu$-type distortion data \cite{Diacoumis:2017hhq,Abitbol:2017vwa}.
\begin{table}
\begin{center}
\begin{tabular}{|c||c|c|c|}
\hline
Parameter (68\%)& free $n_s$ & free $\alpha_s$&free $\beta_s$\\
\hline
$\log(10^{10}A_s)$&$3.1047\pm0.0057$ &$3.1073\pm0.006$  &$3.1061\pm0.0065 $\\
$n_s$&$0.9751\pm0.0045$&$0.973\pm 0.0057$  &$0.9687^{+0.0051}_{-0.006}$\\
$\alpha_s$& N/A&$-0.009\pm0.0067$  &$0.008\pm 0.013$\\
$\beta_s$& N/A & N/A&$0.020\pm 0.013$\\
\hline
\end{tabular}
\end{center}
\caption{Results from 3 analyses of the latest BICEP2+Planck dataset, each adding a free parameter in the power spectrum. The results shown are best fits, within the 68\% confidence level for each analysis. \label{Table_cosmomc}}
\end{table}
\section{Results}
We apply the methods discussed in Section \ref{methods} to the  degree six polynomial inflationary potentials that yield $r=0.01$. We calculate the most likely coefficients and extract the resulting most likely polynomial inflationary potential. The PPS resulting from this inflationary potential is then calculated in order to confirm that the most likely coefficients reconstruct the most likely observables.

\subsection{Results for degree six polynomials that yield $r=0.01$}
\begin{figure}[!ht]
\includegraphics[width=1\textwidth]{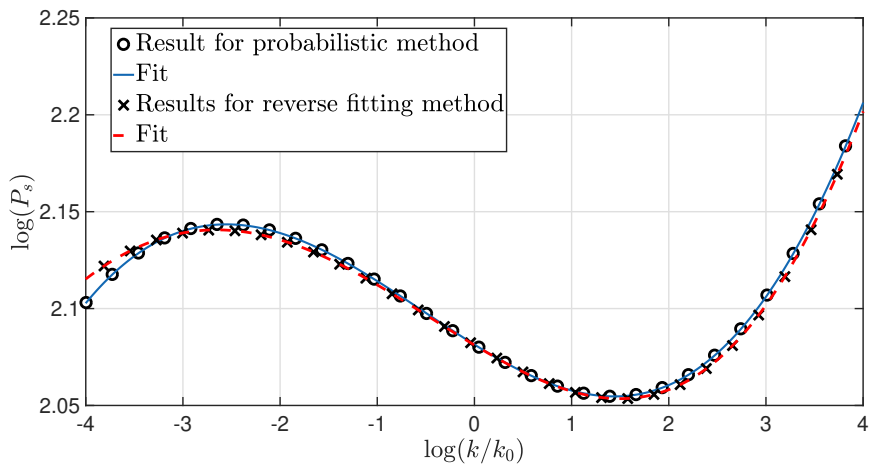}
\caption{Reconstruction of the PPS from the most likely potential with $r=0.01$, as calculated by the multinomial (reverse fitting) method (X's and red dash), as well as the probabilistic method (circles and blue line). The CMB observables are well within the 68\% confidence levels of the MCMC analysis for both. However, the probabilistic method seems to yield more precise results.\label{Reconstruct_001} }
\end{figure}
In Fig. \ref{main_001} we showed a cover for the joint likelihood map of $n_s - \alpha_s$, of about $2000$ potentials with $r=0.01$. The cover is approximately uniform, thus we were able to assign likelihoods to every potential we study, as previously discussed.

By a process of marginalization, as discussed in Section \ref{methods}, we extract the most likely coefficients, which yield the likeliest observables. This process is represented graphically in Fig. \ref{Prob_001}. The results are shown in Table \ref{RES_001} and the PPS reconstruction is shown in Fig \ref{Reconstruct_001}. The advantage of this method is that it also determines the deviation from the average value. This can be used as an indicator for the level of tuning that is required to construct the most likely small field model. A discussion of tuning in field theoretic models can be found in \cite{Barbieri:1987fn}, as well as in \cite{Ellis:1986yg} and \cite{Fowlie:2014xha}. In most cases the tuning level can be viewed as simply $\tfrac{\Delta x_i}{x_i}$, which in this case are given by $(0.375,0.27,5.5)$ for $a_2,a_3,a_4$ respectively. The width of the Gaussian fits for $\{a_2,a_3,a_4\}$ are $\{0.015,0.041,0.112\}$ respectively. These widths represent the effective area in parameter space that yields observables within the $68\%$ CL. Which is another measure of the tuning required in the sixth-order polynomial models.

~ Recalculating the CMB observables that this most likely model yields, we find $n_s=0.9687,\alpha_s=0.0089,\beta_s=0.0176$. These values are very close to the most likely values found from the previously discussed MCMC analysis of the BICEP2+Planck data. The resulting scalar index fits the most likely value in Table \ref{Table_cosmomc} exactly, while $\alpha_s$ and $\beta_s$ deviate from these values by no more than $12.5\%$  . We found that this is a relic of the binning method. Adding more models to the simulated data and refining the binning process results in even better proximity to the desired values.

Using the method of multinomial evaluation (\ref{multi}), we found the multinomial coefficients for each of the model degrees of freedom. For instance for $a_2$ we have:
\begin{align}
	 \begin{array}{ccc}
	B&=&\left(\begin{array}{ccc}
	 -20.97 &-0.936& -37.716\\
	&19.19&30.402\\
	&&-407.53
	\end{array}\right)\\
	A&=&\left(40.918,0.955,79.253\right)\\
	p_0&=&- 19.938
	\end{array} .
\end{align}
Since $n_s \sim 1$, and $\alpha_s$ and $\beta_s$ are of the order of $10^{-2}$, the above result suggests that $a_2$ is primarily dominated by $n_s$. Similarly, we have found that $a_3$ is dominated by a linear combination of $\alpha_s$, and $\beta_s$, and $a_4$ is primarily dominated by $\beta_s$.
This method yields the most likely CMB observables with comparable accuracy to the previous method upon recalculation: $n_s=0.9684,\alpha_s=0.0077,\beta_s=0.020$.
\subsection{Most likely potentials}
\begin{table}[!ht]
\begin{center}
\begin{tabular}{|c||c|c||}
\hline
Observable&\multicolumn{2}{|c|}{Recalculated}\\
\hline
&Probabilistic Method& Multi-fit\\
\hline\hline
$n_s$&$0.9687$&$0.9684$\\
\hline
$\alpha_s$&$0.0089$&$0.0076$\\
\hline
$\beta_s$&$0.0176$&$0.020$ \\
\hline
\end{tabular}
\caption{A comparison of recalculated power spectra observables from results of the two extraction methods.\label{RECALC_001}}
\end{center}
\end{table}
Since $n_s$ is better constrained, we opt for the analysis that yields a more precise value of $n_s$. The leading 6th degree polynomial which yields $r=0.01$ at the proper pivot scale, is thus given by:
\begin{align}
	V=V_0\left(1 - 0.035\phi + 0.04\phi^2 -0.15\phi^3 +0.02\phi^4 +0.76\phi^5 -0.78 \phi^6\right).
\end{align}
Upon initial investigation, it seems these models produce a relatively flat tensor power spectrum. This might motivate future research of the tensor power spectrum predictions and constraints.
\section{Observable dependence}
\begin{table}[!ht]
\begin{center}
\begin{tabular}{|c||c|c|c||c}
\hline
&\multicolumn{2}{|c|}{Gaussian extraction}&Multinomial fit\\
\hline
$r=0.01$&$\mu$ (average)& $\sigma$ (standard deviation)& value\\
\hline\hline
$a_2$&$0.0402$ &$0.0156$&$0.01866$\\
\hline
$a_3$&$-0.152$ &$0.0414$&$-0.0235$\\
\hline
$a_4$&$0.0215$ &$0.1123$&$-0.3452$\\
\hline
\end{tabular}
\caption{The most likely coefficients extracted by the process of likelihood assignment and marginalization, as well as by using the multinomial method. \label{RES_001}}
\end{center}
\end{table}
An interesting finding is an inter-dependence of the three observables $n_s,\alpha_s,\beta_s$. For models that yield $r=0.01$, there is a quadratic relation between the observables, such that $\beta_s=\beta_s(n_s,\alpha_s)$. It should be stressed that this is a phenomenon associated with the models and not with the observational data. This is supported by the small $n_s,\alpha_s,\beta_s$ paired-covariance found in \cite{Cabass:2016ldu}, implying weak dependence among observables in the data itself.
\begin{figure}[h]
\includegraphics[width=1\textwidth]{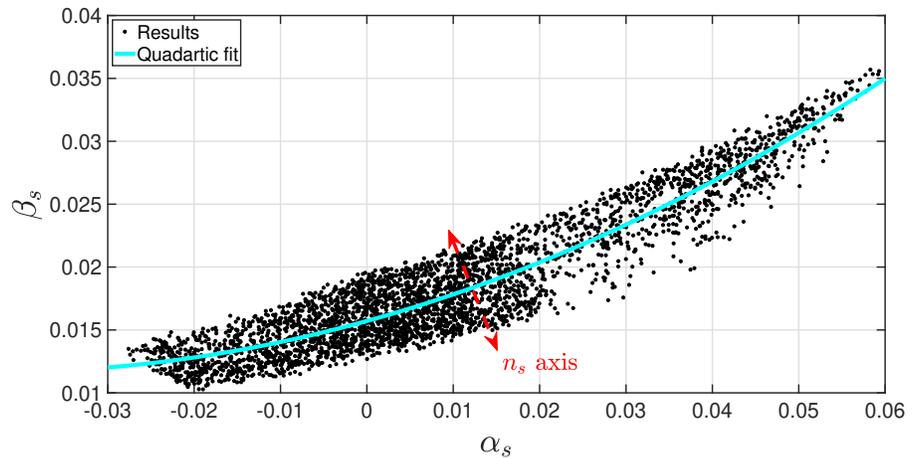}
\caption{Dependence of $\beta_s$ on the other observables, exposes an approximate quadratic relations between $\alpha_s$ and $\beta_s$. The width of the resulting band indicates the deviation from a quadratic relation, which is correlated to $n_s$.
\label{Dependency2} }
\end{figure}

 One might think that the previous findings in \cite[Figure~23]{Ade:2015xua} indicate that $n_s$ and $\alpha_s$ are dependent. However, the graph shows a dependence between $\alpha_s$ and $n_{s,0.002}$ which is the scalar index evaluated at $k=0.002\; h\;Mpc^{-1}$. Taking some initial $n_s$ evaluated at $k_{0}$, it follows that $n_s$ evaluated at some other scale, depends on the index running $\alpha_s$. Indeed, when one examines the color coding in \cite[Figure~23]{Ade:2015xua}, which represents $n_s$ at the pivot scale, it is clear that $n_s$ and $\alpha_s$ are independent.

\section{Summary and outlook}
A large sample of potentials that yield $r=0.01$ and conform to the allowed observable values was successfully generated. The sample provided a uniform cover of the allowed region of parameters which enabled us to assign likelihoods to each of the potentials and extract each coefficient's likelihood (\ref{Probability}). Another approach was implemented, representing each coefficient as a multinomial function of the observables (\ref{multi}), which yielded similar results. A most likely small field potential giving rise to $r=0.01$ and $(n_s\simeq 0.9694,\alpha_s\simeq 0.009,\beta_s\simeq 0.0175)$, was identified, and its power spectrum simulated. An interesting inter-dependence of $(n_s,\alpha_s,\beta_s)$ was found in these models, which may have some bearing on future MCMC analysis. We hope to perform such an MCMC analysis, either with priors that include this dependence or with a numerical scheme that reflects it.

~ The Planck collaboration may soon release additional analysis products, and BICEP3 is also expected to release results in the near future. Thus, it might be possible to check our prediction for the tensor-to-scalar ratio. However, ruling out models of the class discussed in this paper might be a more difficult task due to the lack of constraining power of current observations in the range $l>1500$. We expect that either S4 cosmology or $\mu$-distortion data will be able to resolve this in the foreseeable future by adding observational data on smaller scales.

\section*{Acknowledgements}
The research of RB and IW was supported by the Israel Science Foundation grant no.
1294/16.

\end{document}